\documentstyle[prl,preprint,aps,prabib,epsf]{revtex}

\begin{document}
\draft

\title{Folding Transitions of Self-Avoiding Membranes}

\author{Christian M\"unkel\cite{muenkel} and Dieter W. Heermann}
\address{
	Institut f\"ur Theoretische Physik \\
	Universit\"at Heidelberg \\
	Philosophenweg 19\\
	69120 Heidelberg\\
and\\
	Interdisciplinary Center for Scientific Computing\\
	University of Heidelberg\\
	Im Neuenheimer Feld 368\\
	69120 Heidelberg\\
Germany
     }
\date{accepted by Physical Review Letters}

\maketitle

\begin{abstract}
{\bf Abstract:}
The phase structure of self-avoiding polymerized membranes
is studied by extensive {{\em Hybrid Monte Carlo}\ }
simulations. Several folding transitions from the flat to a collapsed state
are found.
Using a suitable order parameter and finite size scaling theory,
these transitions are shown to be of {\em first order}.
The phase diagram in the temperature-field plane is given.
\end{abstract}



Fluctuating membranes and surfaces are basic structural elements of
biological systems and complex fluids. Recent theoretical work
\cite{nelson89,abraham90a} and experimental studies \cite{spector94,mutz91}
 indicate, that these sheetlike macromolecules should have
dramatically different properties than linear polymers.
Polymerized membranes which contain a permanently cross-linked
network of constituent molecules have a shear elasticity, giving them a large
entropic bending rigidity.
In the absence of self-avoidance, polymerized \cite{kantor86}
and fluid \cite{muenkel93} membranes adopt a crumpled random structure.
Theoretical predictions by Flory mean-field approximation and
Monte Carlo simulations \cite{kantor86} and renormalization group studies
\cite{rg} supported the existence of a high temperature crumpled phase for
self-avoiding polymerized membranes also, suggesting  a
possible finite temperature crumpling transition in the presence of
an explicit bending rigidity \cite{kantor87}.
However, more extensive computer simulations
\cite{abraham90a,abraham89,ho89,plischke88}
 found no crumpling of self-avoiding tethered membranes in a good
solvent. This prediction was confirmed recently by experimental
studies of graphitic oxide \cite{spector94}.

On the other hand, polymerized vesicles undergo a
wrinkling transition \cite{mutz91}, and upon addition of
10 vol \% acetone, Spector and co-workers \cite{spector94} found
small compact objects, which appeared to be folded.
A poor solvent leads to (short-ranged) attractive interactions,
and a single membrane was found to be flat for high
temperatures \cite{abraham89}, but in a collapsed state for sufficiently
low temperatures \cite{abraham90a}. The transition between
the flat and the collapsed states of the membrane proceeds through a
sequence of {\em folding transitions}, which were first
found by cooling of a single membrane
from the flat phase \cite{abraham91}. Because no hysteresis was found,
it was ruled out
that the folded configurations are metastable states.
However, this method does {\em not}
 give sufficient evidence of the
order or even the existence of a transition. For instance, hysteresis
can also be found at {\em second order} phase transitions of finite
systems and
the results for one system size may be misleading.
In addition, the experimentally observed wrinkling transition is first order
\cite{mutz91}.

On that account, we present
 a systematic finite size scaling analysis of the
folding transitions. The {{\em Hybrid Monte Carlo}\ }
 algorithm was used, which provides for
simulations of the canonical ensemble with constant temperature.
Moreover, ensemble averages are independent of the discretization step
size ${\delta}t$,
i.e. systematic errors \cite{mehlig92,duane87}.
A suitable order parameter is defined and a phase diagram is presented.

Besides the nearest neighbour interactions, the membranes were modeled
similar to those of Abraham and Kardar \cite{abraham91}.
The $N$ particles  of the polymerized membrane form
the sites of a hexagonal shaped triangular lattice.
The bond potential between nearest neighbour particles is
\begin{equation}
V^{\mbox{{\tiny B}}} = \sum_{j(i)} {\left( {b_0 - r_{ij}} \right)}^2 \quad ,
\end{equation}
with an equilibrium length $b_0$ and distances $r_{ij}$ between particle
$i$ and its nearest neighbors $j(i)$. In place of this harmonic potential,
tethers were used by Abraham and Kardar \cite{abraham91}.

All particles interact through a shifted Lennard-Jones potential
\begin{equation}
V^{\mbox{{\tiny LJ}}}_{ij} = \left\{
\begin{array}{ccc}
\left( { \frac{1}{r_{ij}^{12}} - 2\frac{1}{r_{ij}^{6}} } \right) +
\left( { \frac{1}{r_{c}^{12}} - 2\frac{1}{r_{c}^{6}} } \right)
& , & r_{ij} <= r_c\\
0 & , & r_c < r_{ij}
\end{array}
\right.
\end{equation}
with a cut-off at $r_c~=~2.5$ \quad . The repulsive part of this
interaction guarantees self-avoidance of the membrane.

The folding of the membrane can be described by the eigenvalues
${\lambda_1}^2~>=~{\lambda_2}^2~>=~{\lambda_3}^2$ of the inertia tensor
\begin{equation}
{\cal T}_{\alpha,\beta} = \frac{1}{N} \sum_{i = 1}^{N}
\left( { r_{i\alpha}r_{i\beta} - \overline{r}_\alpha \overline{r}_\beta }
\right) \quad ,
\label{equ:tensor}
\end{equation}
where $\alpha,\beta \in \{x,y,z\}$, and the sum runs over all particles
of a given configuration; $\overline{r}_\alpha$ is the $\alpha$
component of the center of mass for a configuration.
We can estimate the expected change of the membrane eigenvalues
by those of an unfolded or folded disc with radius $1$ and
width $d=0.1$ as shown in Table~\ref{tab:eig_disc}.
Taking into account all three eigenvalues, an unfolded, folded or
twice folded configuration can be distinguished.
A collapsed configuration would be indicated by
approximately equal eigenvalues.

Membranes with up to $N=1141$ particles were simulated at
inverse temperatures in the range $\beta\in[0.2;0.75]$.
Configurations at $\beta=0.20$, $0.27$ and $0.40$ were found to be
in an unfolded, folded and twice folded state by visualization
(figure not shown).

\begin{figure*}[p]
\epsfysize=9.5cm \epsfbox{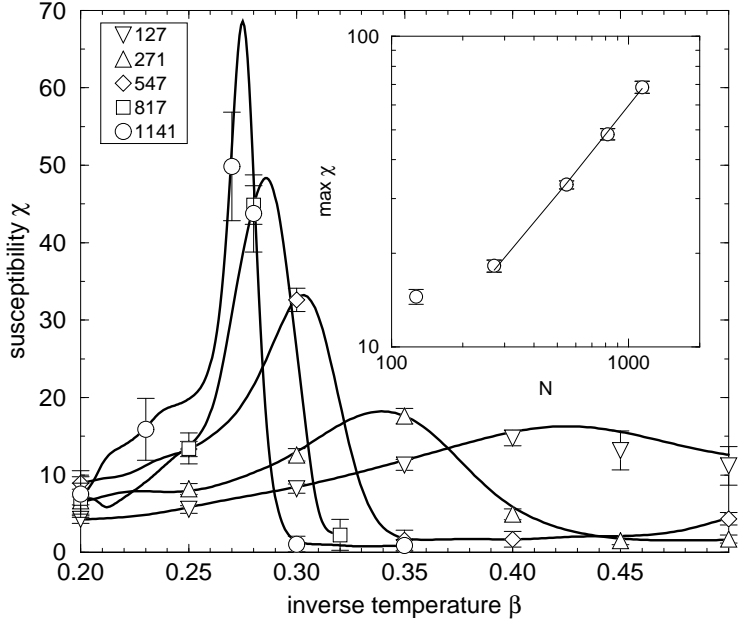}
 \caption{Susceptibility ${\chi}(\beta)$ for membranes with
$N=127,271,547,817$ and $1141$ particles. The solid lines were computed by
the multi-histogram method. The scaling of the
maximum $\max{\chi}$ is shown in the inset.
 }
\label{fig:memb_chi}
\end{figure*}

At the first folding transition, the eigenvalue $\lambda_1^2$
 stays constant approximately, while the second eigenvalue $\lambda_2^2$
decreases
by a factor $c~{\approx}~0.2790$. Therefore, we define an {\em order parameter}
$m$ by
\begin{equation}
\langle{m(\beta)}\rangle = \left\langle {
\frac{1}{\left({1-\sqrt{c}}\right)}
 \left[ {
 \frac{\lambda_2(\beta)}{\lambda_2(\beta_0)} - \sqrt{c}
 } \right]
}\right\rangle \quad,
\label{equ:def_opa2}
\end{equation}
where ${\lambda_2(\beta_0)}$ is a reference value of an unfolded membrane
at $\beta_0$ far below the inverse critical temperature $\beta_c$.
At a first order transition,
the average order-parameter discontinuously jumps at $\beta_c$
\cite{landau88,binder86,binder84}.
Contrary, at a continuous phase transition
we expect a power law behavior
of the average order parameter $\langle{m}\rangle(\beta)$
 with an exponent $\beta^\prime$, i.e. \(
\langle{m}\rangle \propto {\left({\beta - \beta_c}\right)}^{\beta^\prime} ,
 \beta > \beta_c
\) \cite{privman90,barber83,brezin82,fisher72}.
Indeed, the average order parameter
$\langle{m}\rangle$ becomes very steep near $\beta_c$ with increasing
number of particles $N$ (data not shown).
In principle, one can measure the increase of the slope and compare to the
predictions of finite size scaling theory.
Of course, the slope of $\langle{m}\rangle(\beta)$ is proportional to the
{\em susceptibility} $\chi$, which can be measured by the fluctuations
of $m$ also:
\begin{equation}
\chi(\beta) = L^d \beta
\left( { {\langle{m^2}\rangle}_L - {\langle{m}\rangle}_L^2 } \right)
\label{equ:def_chi}
\end{equation}
The susceptibility $\chi(\beta)$ is shown in Figure~\ref{fig:memb_chi} and
the scaling of the maximum $\max{\chi(\beta)}$ in the inset of
Figure~\ref{fig:memb_chi}.

\begin{figure*}[p]
\epsfysize=9.5cm \epsfbox{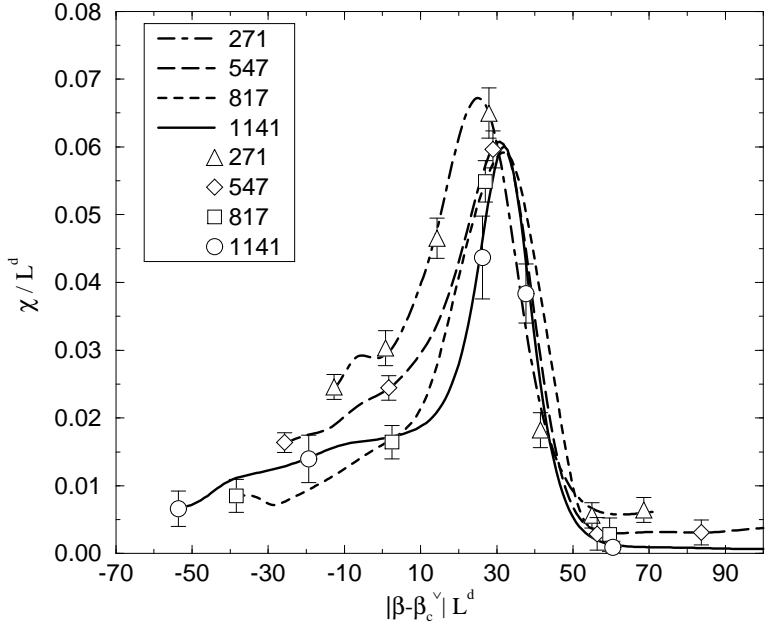}
 \caption{Finite size scaling plot of the susceptibility $\chi(\beta)$
for a first order transition using the value $\beta_c^\infty=0.247(5)$
 from Figure~\protect\ref{fig:memb_beta_c}. }
\label{fig:memb_chi_fss}
\end{figure*}

At a first order transition, $\chi(\beta_c^{\mbox{\tiny eff}})$
is expected to increase proportional to
$N=L^d$~\cite{landau88,binder86,binder84}.
At a continuous phase transition,
${\left|{\beta-\beta_c}\right|}^{-\gamma_\pm}$
is predicted \cite{privman90,barber83,brezin82,fisher72}.
The $L^d$ increase of $\chi(\beta)$
in Figure~\ref{fig:memb_chi} gives evidence of the scaling
at a {\em first order} transition.
In addition, the width of the susceptibility peak should
decrease as $L^{-d}$ \cite{landau88,binder86,binder84}.
Figure~\ref{fig:memb_chi_fss} shows a finite size scaling plot of the
susceptibility data. Within errors, {\em first order} scaling behavior can be
observed, at least above the inverse transition temperature.

\begin{figure*}[p]
\epsfysize=9.5cm \epsfbox{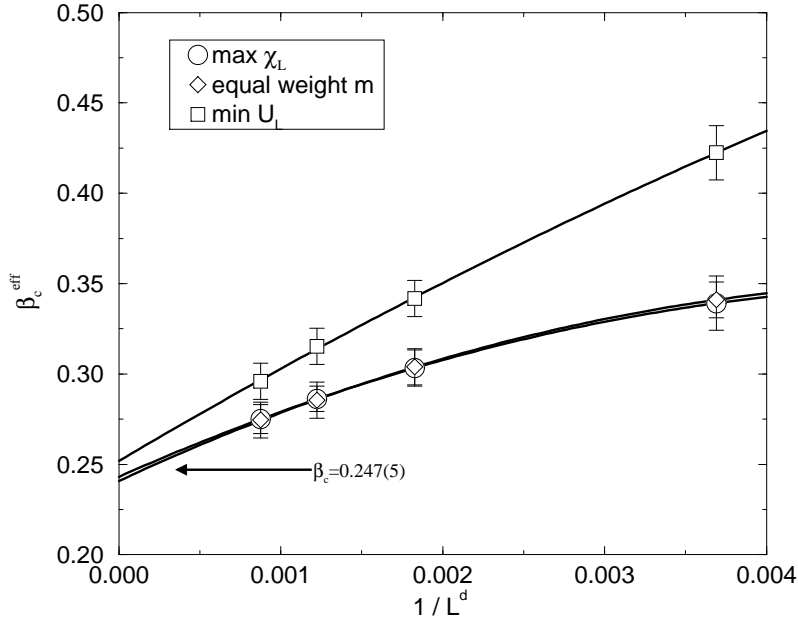}
 \caption{Critical temperature $\beta_c^\infty=0.247(5)$ determined by the
position of the maximal susceptibility $\chi$, the minimum of the
cumulant $U_L$ and equal-weight of the order parameter distribution
for the four largest system sizes.
 }
\label{fig:memb_beta_c}
\end{figure*}

The transition temperature $\beta_c^\infty$ can be extrapolated
by the position of the maximum in $\chi(\beta)$ \cite{binder84},
 the minimum of the
cumulant \cite{binder84} and the equal weight criterion \cite{borgs92}
 of the order parameter
distribution. At least for $\chi$, the equal weight criterion predicts
a shift of the effective transition temperature proportional
to $L^{-2d}$ \cite{borgs92a,borgs92,borgs90}, whereas a shift
proportional to $\propto L^{-1/\nu}$ is expected for a continuous
phase transition.
 Using quadratic terms in the regression, the extrapolations
in Figure~\ref{fig:memb_beta_c}
of the three observables agree within errors.
The transition temperature of the first folding transition is found to be
$\beta_c^\infty=0.247(5)$.

\begin{figure*}[p]
\epsfysize=9.5cm \epsfbox{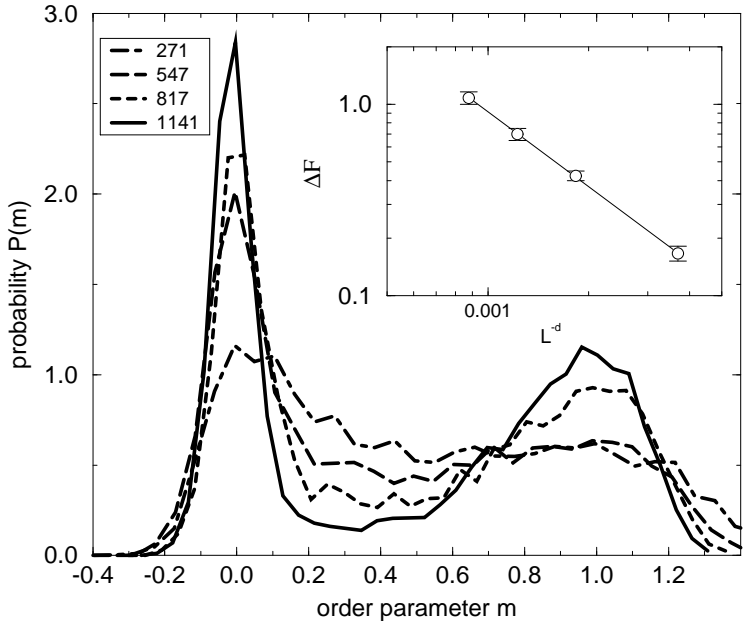}
 \caption{Probability distribution $P(m,\beta_{ew})$ of the order parameter
$m$ at the equal-weight transition temperature $\beta_{ew}$. The inset
shows the free energy difference ${\Delta}F$ at the equal-height
transition temperature $\beta_{eh}$, which increases
$\propto {(L^d)}^x$, $x{\approx}1.3$.
 }
\label{fig:memb_ew_eh}
\end{figure*}

The increase of the susceptibility is caused by
the characteristic double-peak structure of the order parameter distribution
near $\beta_c$, which
is typical for a discontinuous phase transition
\cite{landau88,binder86,binder84,privman90,borgs92a,borgs92,borgs90,kosterlitz90}.
Figure~\ref{fig:memb_ew_eh} shows the expected double-peak
distribution $P(m)$ at the equal height transition temperature
\cite{borgs92}. The development of a minimum in $P(m)$ is confirmed by
the method of Lee and Kosterlitz \cite{kosterlitz90}. The measured
${\Delta}F$ in the inset of Figure~\ref{fig:memb_ew_eh} is
proportional to the free energy difference at the equal height transition
temperature and increases as ${(L^d)}^x$, $x{\approx}1.3$.

Further, $P(m)$ can be described by the reduced cumulant $U_L$
\begin{equation}
U_L(\beta) = 1 - \frac{{\langle{m^4}\rangle}_L}{3{\langle{m^2}\rangle}^2_L}
\quad .
\label{equ:def_U_L}
\end{equation}
At a continuous phase transition, $U_L(\beta)$ is expected to approach
$2/3$ for all $\beta$. The data shows
a minimum, which becomes more pronounced for large $N$, indicating
a {\em first order} phase transition (figure not shown).

\begin{figure*}[p]
\epsfysize=9.5cm \epsfbox{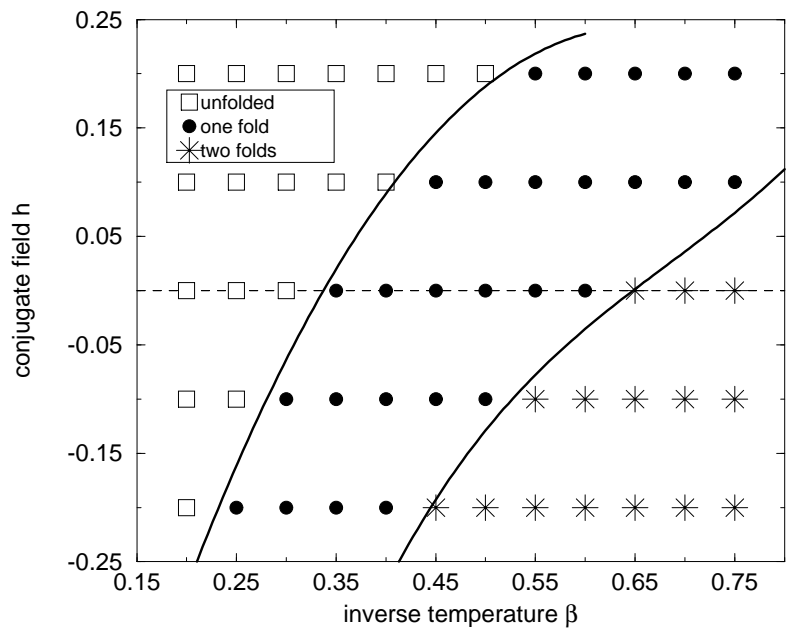}
 \caption{Phase diagram of a membrane with $271$ particles in
the $\beta,h$ plane. The symbols denote the result of the simulations,
the solid lines are the first order transition lines computed by the
multi-histogram method.
 }
\label{fig:memb_phase}
\end{figure*}

The folding of the membrane must be visible in the
attractive part of the potential energy, also. In fact, there is a jump
in the potential energy and the related specific heat develops a
peak, although very slowly.
Besides the above defined order parameter, which is based on the
geometry of the membrane, we can derive a different order parameter
from the attractive part of the Lennard-Jones potential:
\begin{equation}
\tilde{m} = \frac{1}{N} \left\{
	\sum_{i<j} \Theta \left( {r_c - r_{ij}} \right)
	\left( { 2\frac{1}{r_{ij}^{6}} + 2\frac{1}{r_{c}^{6}} } \right)
\right\} \quad .
\label{equ:def_opa1}
\end{equation}
Compared to $m$, the order parameter $\tilde{m}$ has the advantage, that
it is a local property. We performed {{\em Hybrid Monte Carlo}\ }
 simulations with an additional term
$h{\cdot}\tilde{m}$ in the Hamiltonian, where $h$ is the conjugate field.
Figure~\ref{fig:memb_phase} shows the phase diagram in the $(\beta,h)$-plane
of a membrane with $271$ particles. The first order transition lines
were computed by the multi-histogram method \cite{ferrenberg89}.
For the transition from one to two folds, an order parameter similar
to $m$ is used, which is based on $\lambda_1$ instead of
$\lambda_2$.
$h=1$ is an upper limit for both transition lines, because of the
vanishing attractive interactions at $h=1$.

The main result of this study is the {\em first order} nature of the
folding transitions of self-avoiding polymerized membranes.
This is in agreement with the observed folded structure of
graphitic oxide \cite{spector94} and the first order wrinkling transition
of partially polymerized vesicles \cite{mutz91}.

Recently, another first order folding transition was found by
DiFrancesco and Guitter \cite{guitter94} for
regular triangular lattices embedded into two-dimensional space.
However, these transitions combine flat and compact states directly, i.e.
without several distinct foldings. Therefore, the folding
transition described here is of a different kind.

The unfolding of a singly folded membrane bears close resemblance to the
unbinding transition of two distinct surfaces. Regarding a folded
membrane of $N$ particles,
the fraction of particles near the crease decreases with $1/\sqrt{N}$.
Therefore, the nature of the folding transition is related to the
nature of the
underlying unbinding transition of two distinct membranes without the
crease. The shape fluctuations of a single membrane of
lateral size $\xi_\parallel$
are characterized by the typical fluctuation amplitude
$\xi_\perp \propto {\xi_\parallel}^{\zeta}$.
Polymerized membranes without lateral tension have
a roughness exponent $\zeta\approx0.6$.
The steric hindrance of two interacting membranes at separation $l$
leads to an overall loss of entropy, which can be regarded
as an effective fluctuation-induced repulsion,
$V_{rep}\propto{1/{l^\tau}}$ with decay exponent $\tau\approx3.3$
for polymerized membranes. This repulsive interaction causes
the unfolding of the membrane even in the presence of attractive
van der Waals interactions.
However, the crease of the folded membrane introduces an additional
attractive interaction. This situation is similar to a
membrane interaction which exhibits two minima at two different separations.
Such an interaction implies a {\em first-order} unbinding transition
\cite{lipowsky90} and may be an explanation of
the {\em first-order} folding transition described in
this work.

Further investigations are
necessary to determine the nature of the relation between folding and
unbinding of polymerized membranes.

\acknowledgments
Funding of this work by BMFT project 031240284 and LGFG 9127.1
 is gratefully acknowledged.
Some of the simulations were done at the {\em Interdisziplin\"ares
Zentrum f\"ur wissenschaftliches Rechnen} of the University Heidelberg.


\tighten
\onecolumn

\begin{table}
\begin{center}
\begin{tabular} {|l|ccc|}  \hline
configuration & ${\lambda_1}^2$ & ${\lambda_2}^2$ & ${\lambda_3}^2$ \\
\hline
disc & $0.2543$ & $0.2473$ & $0.00084$ \\
disc, folded & $0.2524$ & $0.0690$ & $0.00332$ \\
disc, folded twice & $0.0929$ & $0.0487$ & $0.01330$ \\
\hline
disc & $1.0000$ & $1.0000$ & $1.00$ \\
disc, folded & $0.9925$ & $0.2790$ & $3.95$ \\
disc, folded twice & $0.3653$ & $0.1969$ & $15.83$ \\
\hline
\end{tabular}
\end{center}
\caption{Eigenvalues of the moment of inertia tensor of a disc with
radius $1$ and width $d=0.1$. In the lower part, all numbers are relative
to the values of the unfolded disc.
}
\label{tab:eig_disc}
\end{table}

\eject

\newpage

\end{document}